\documentclass[journal]{IEEEtran}

\usepackage{graphicx,amsmath,amssymb,amsfonts,cite}
\usepackage{subfigure}
\usepackage{algorithm}
\usepackage{algorithmic}
\usepackage{float}
\usepackage{amsthm}
\usepackage{ulem}
\usepackage{bm}
\usepackage{subfigure}
\usepackage{multirow}
\usepackage{array}
\usepackage{color}
\usepackage{caption}

\newcolumntype{I}{!{\vrule width 3pt}}
\newlength\savedwidth

\newlength\savewidth

\ifCLASSINFOpdf
\else
\fi
\begin{document}
%
\title{Multiple Preambles for High Success Rate of Grant-Free Random Access with Massive MIMO}
%
%
%

\author{Hao~Jiang,~
        Daiming~Qu,~ 
        Jie~Ding, and~Tao~Jiang,~\IEEEmembership{Senior Member,~IEEE}
\thanks{Hao Jiang, Daiming Qu, Jie~Ding and Tao Jiang are with
School of Electronic Information
and Communications, Huazhong University of Science and Technology,
Wuhan, 430074, China.}
\thanks{This work was supported in part by the National Natural Science Foundation of China funded project under grant number 61701186 and 61571200, and in part by the China Postdoctoral Science Foundation funded project under grant number 2017M612458.}
}

\maketitle

\begin{abstract}
Grant-free random access (RA) with massive MIMO is a promising RA technique with low signaling overhead that provides significant benefits in increasing the channel reuse efficiency. Since user equipment (UE) detection and channel estimation in grant-free RA rely solely on the received preambles, preamble designs that enable high success rate of UE detection and channel estimation are very much in need to ensure the performance gain of grant-free RA with massive MIMO. In this paper, a super preamble consisting of multiple consecutive preambles is proposed for high success rate of grant-free RA with massive MIMO. With the proposed approach, the success of UE detection and channel estimation for a RA UE depends on two conditions: 1) it is a solvable UE; 2) its super preamble is detected. Accordingly, we theoretically analyze the solvable rate of RA UEs with multiple preambles and propose a reliable UE detection algorithm to obtain the super preambles of RA UEs by exploiting the quasi-orthogonality characteristic of massive MIMO. Theoretical analysis and simulation results show that turning a preamble into a super preamble consisting of two or three shorter preambles, the success rate of UE detection and channel estimation could be significantly increased using the proposed approach.
\end{abstract}


\begin{IEEEkeywords}
Random access, Grant-free, Preamble design, Massive MIMO, M2M.
\end{IEEEkeywords}

%
\IEEEpeerreviewmaketitle

\section{Introduction}
%
%
%
%
\IEEEPARstart{F}uture wireless networks are expected to accommodating a rapidly growing number of connected devices and handling their respective data traffic, such as the Internet of Things (loT) \cite{Stankovic2014} \cite{Schulz2017}. As an important enabler of the loT, machine-to-machine (M2M) communications have attracted considerable attention from academia and industries. M2M communications are commonly characterized by a massive number of intermittent active user equipments (UE) with small-sized data payloads. To fulfilling the demand of massive access, massive MIMO, which is a promising technique to greatly increase capacity for future wireless communications \cite{Marzetta2010,Larsson2014,Lu2014}, is being considered to support M2M communications\cite{Carvalho2017,Carvalho2017TWC,Bjornson2016,Han2017}. However, considering small data payloads, the conventional request-grant random access (RA) procedure in Long Term Evolution (LTE) is not efficient due to the significant signaling overhead \cite{Biral2015,Zanella2013,Madueno2014,Laya2014}. To minimize signaling overhead, grant-free RA protocols, where UEs contend (i.e, perform random access) directly with their uplink data payloads by transmitting preamble along with data, is being considered as an alternative for M2M communications \cite{Mahmood2016}. As a result, the radio resources reserved in the request-grant procedure could be unleashed for accommodating more RA UEs. With massive MIMO, the radio resources saved by grant-free could be used for accommodating more UEs compared to single-antenna systems. Therefore, grant-free RA with massive MIMO is being considered as a compelling alternative for M2M communications\cite{Ding2018}.

Our previous works in \cite{Ding2018} confirmed that massive MIMO provides significant benefits for grant-free RA in increasing the channel reuse efficiency, which enables multiple UEs to access a single channel with grant-free data transmission. It is also found that the performance of grant-free RA with massive MIMO is mainly dominated by the number of orthogonal preambles as long as the number of antennas is sufficiently large \cite{Ding2018}. The reason is that UE detection and channel estimation in grant-free RA rely solely on the received preambles, which is very different from the case of request-grant RA, where UE detection could rely on the contention resolution mechanism through protocol exchange and channel estimation is carried out at the latter non-contention stage. For instance, when preamble collision occurs in grant-free RA, i.e., multiple RA UEs select the same preamble, the base station (BS) can only detect one RA UE from this preamble and its channel response would be incorrectly estimated. Consequently, the data payloads of the RA UEs that involved in the preamble collision are unlikely to be recovered. Moreover, the incorrect channel responses lead to an incorrect beamforming pattern (especially for zero-forcing beamforming), which would bring in multiuser interference to other RA UEs and degrade the error performance for all RA UEs as a result. Therefore, preamble designs that enable UE detection and channel estimation with high success rate, to support recovery of the following data packet, are very much in need to ensure the performance gain of the grant-free RA with massive MIMO.

In this paper, we propose a multiple-preamble grant-free RA approach with massive MIMO. In the uplink phase, each RA UE transmits a \emph{super preamble}, which consists of $L$ consecutive preambles, followed by a data payload. The BS relies on the received super preambles to detect the transmitting UEs, make channel estimations, and then recover the data payloads. In each preamble phase, each RA UE randomly selects a preamble sequence among a common preamble sequence pool. To represent the super preambles that the RA UEs select, a matrix with elements of zero and one is formulated and referred to as \emph{preamble selection matrix}. The UE detection under the proposed approach is to obtain this preamble selection matrix, which is equivalent to detecting the super preambles transmitted by the RA UEs. To fulfill this objective, we propose a reliable UE detection algorithm by exploiting the quasi-orthogonality characteristic of massive MIMO. After the UE detection, channel estimation can be easily obtained by matrix operations involving the inverse of the preamble selection matrix. We demonstrate that the probability that the preamble selection matrix is full row rank is high with the proposed multiple-preamble approach, thus the channel responses of the RA UEs can be acquired with high success rate. Theoretical analysis and simulation results show that turning a preamble into a super preamble consisting of two or three shorter preambles, the success rate of UE detection and channel estimation could be significantly increased using the proposed approach.

The multiple-preamble structure in the proposed approach is inspired by Code-expanded Random Access (CeRA) \cite{Pratas2012}. CeRA is a kind of request-grant RA protocol, where each RA UE transmits a sequence of preambles as a codeword, called as super preamble in this paper, instead of a single preamble to request the access to the uplink radio resources. The BS detects preambles at each preamble phase and take all combinations of detected preambles as possible codewords the RA UEs sent. Then, the BS sends a number of RA responses, each of which corresponds to a possible codeword and a granted resource for uplink data transmission. As a result, CeRA provides a significant increase in the amount of available contention resources, and enables the service of an increased number of RA UEs. Different from CeRA, our aim is to make UE detection and channel estimation directly from the received super preambles with high success rate so that grant-free RA with high performance gain is supported, without additional protocol exchange.

The remainder of this paper is organized as follows. In Section II, the multiple-preamble grant-free RA with massive MIMO is briefly described.
In Section III, analysis on the solvable rate of RA UE (UEs) are detailed. In Section IV, the UE detection algorithm with the support of massive MIMO is proposed. Simulation results are presented in Section V and the paper is concluded in Section VI.

\emph{Notations:} Boldface lower and upper case symbols represent
vectors and matrices, respectively. $\mathbf{I}_n$ is the $n \times n$ identity matrix. The $i$th row, the $j$th column and the $i$th row and the $j$th
column element of a matrix $\mathbf{X}$ are denoted by $(\mathbf{X})_{i,-}$, $(\mathbf{X})_{-,j}$ and $(\mathbf{X})_{i,j}$, respectively. The transpose, conjugate-transpose and the Moore-Penrose inverse of a matrix $\mathbf{X}$ are denoted by $\mathbf{X}^{T}$, $\mathbf{X}^{H}$ and $\mathbf{X}^{+}$, respectively. The modulus of a complex-valued number $x$ is denoted as $|x|$ and the Euclidean norm of a vector $\mathbf{x}$ is denoted as $||\mathbf{x}||$. We use $\mathbb{C}$ to denote spaces of complex-valued numbers. $\mathbf{x}\sim \mathcal{CN}(0,\mathbf{\Sigma})$ indicates that $\mathbf{x}$ is a symmetric complex Gaussian random vector with zero-mean and covariance matrix $\mathbf{\Sigma}$.
\hfill

\section{Multiple-Preamble Grant-Free RA Model}
\begin{figure}[!t]
\centering
\includegraphics[width=3.5in]{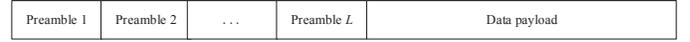}
\caption {The frame structure of multiple-preamble grant-free RA} \label{fig11}
\end{figure}

We consider a single cell massive MIMO network consisting of an $M$-antenna BS and $N$ single-antenna UEs, where the $N$ UEs are attempting random access simultaneously over a same channel. As depicted in Fig. \ref{fig11}, each RA UE transmits a super preamble, which consists of $L$ consecutive preambles, followed by a data payload. In each preamble phase, each UE randomly selects a preamble sequence from a common preamble sequence pool consisting of $K$ orthogonal preamble sequences. The preamble sequence pool is denoted as $\mathbf{S} =[\mathbf{s}_{1}, \mathbf{s}_{2}, . . . , \mathbf{s}_{K}]^{T}\in\mathbb{C}^{K \times K}$, satisfying $\mathbf{S}\mathbf{S}^{H} = \mathbf{I}_{K}$, where $\mathbf{s}_{k}$ ($k=1,2,...,K$) is a preamble sequence of $K$ symbols. We assume the duration of the uplink RA frame is smaller than the channel coherence interval so that the channel between each RA UE and the BS could be described by a constant channel response within a frame. We also assume the power control is applied to keep the received power at the BS from all RA UEs at approximately the same level. Therefore, the preamble signal received at the $M$-antenna BS in preamble phase $l$ ($l=1,2,...,L$), denoted by $\mathbf{Y}_{l}\in\mathbb{C}^{M \times K}$, is given by
\begin{align}\label{eq1}
\mathbf{Y}_{l}=\mathbf{H}\mathbf{P}_{l}+\mathbf{N}_{l},
\end{align}
where $(\mathbf{Y}_{l})_{m,k}$ is the $k$-th ($k=1,2,...K$) sample at the $m$-th ($m=1,2,...M$) antenna in preamble phase $l$, $\mathbf{H} = [\mathbf{h}_{1},..., \mathbf{h}_{N}]\in\mathbb{C}^{M \times N}$ represents the uplink channel response matrix from the $N$ RA UEs to the BS, i.e., $\mathbf{h}_{n}= [h_{n,1},h_{n,2},..., h_{n,M}]^{T}$ ($n=1,2,...N$) is the channel response vector between UE $n$ and the BS, where $h_{n,m}$ is the channel response between UE $n$ and the $m$-th antenna of the BS, $\mathbf{P}_{l}= [\mathbf{p}_{1,l},\mathbf{p}_{2,l},..., \mathbf{p}_{N,l}]^{T}\in\mathbb{C}^{N \times K}$ is the preamble sequence matrix transmitted by all the RA UEs in preamble phase $l$, i.e., $\mathbf{p}_{n,l}^{T}$ is a row vector representing the preamble sequence transmitted by UE $n$ and it is equivalent to one of the row vectors of $\mathbf{S}$, $\mathbf{N}_{l}$ is the complex additive white Gaussian noise matrix at the BS.

After the BS receives the preamble signal $\mathbf{Y}_{l}$, it correlates $\mathbf{Y}_{l}$ with $\mathbf{S}$. The correlation result, denote by $\mathbf{B}_{l}\in\mathbb{C}^{M \times K}$, is given by
\begin{align}\label{eq2}
\mathbf{B}_{l} &=\mathbf{Y}_{l}\mathbf{S}^{H} \nonumber\\
&=\mathbf{H}\mathbf{P}_{l}\mathbf{S}^{H} + \mathbf{N}_{l}\mathbf{S}^{H}.
\end{align}
Let $\mathbf{A}_{l} = \mathbf{P}_{l}\mathbf{S}^{H}$, where $\mathbf{A}_{l}\in\mathbb{C}^{N \times K}$ and its elements are either zero or one. A row vector of $\mathbf{A}_{l}$ indicates the preamble sequence selected by the corresponding RA UE in the preamble phase $l$, i.e., if the $n$-th row and $k$-th column element of $\mathbf{A}_{l}$ equals to one, it indicates that UE $n$ transmits $\mathbf{s}_{k}$ in preamble phase $l$.
Then, $\mathbf{B}_{l}$ is rewritten as
\begin{align}\label{eq3}
\mathbf{B}_{l}&=\mathbf{H}\mathbf{A}_{l} + \mathbf{W}_{l},
\end{align}
where $\mathbf{W}_{l}=\mathbf{N}_{l}\mathbf{S}^{H}$.

\begin{figure}[!t]
\centering
\includegraphics[width=3.5in]{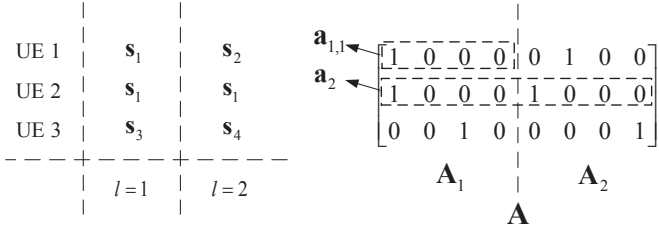}
\caption {Preamble selection matrix with $K=4$, $L=2$, $N=3$.} \label{fig13}
\end{figure}

Considering all the preamble phases, the correlation results of the $L$ preamble signals with $\mathbf{S}$ is given by
\begin{align}\label{eq4}
\mathbf{B} =\mathbf{H}\mathbf{A}+\mathbf{W},
\end{align}
where $\mathbf{B} = [\mathbf{B}_{1},\mathbf{B}_{2},...,\mathbf{B}_{L}]\in\mathbb{C}^{M \times KL}$, $\mathbf{A} = [\mathbf{A}_{1},\mathbf{A}_{2},...,\mathbf{A}_{L}]\in\mathbb{C}^{N \times KL}$ and $\mathbf{W} = [\mathbf{W}_{1},\mathbf{W}_{2},...,\mathbf{W}_{L}]\in\mathbb{C}^{M \times KL}$. $\mathbf{A}$ is referred to as preamble selection matrix and it could also be written as $\mathbf{A}=[\mathbf{a}_{1}^{T}, \mathbf{a}_{2}^{T}, . . . , \mathbf{a}_{N}^{T}]^{T}$, where $\mathbf{a}_{n}\in\mathbb{C}^{1\times KL}$ ($n=1,2,...,N$) is referred to as \emph{preamble selection vector}. Vector $\mathbf{a}_{n}$ consists of $L$ sub-vectors with length of $K$, i.e., $\mathbf{a}_{n} =[\mathbf{a}_{n,1},\mathbf{a}_{n,2},...,\mathbf{a}_{n,L}]$, where $\mathbf{a}_{n,l} \in\mathbb{C}^{1\times K}$ and each of the $L$ sub-vectors indicates the preamble sequence selected by UE $n$, i.e., if the $k$-th column of $\mathbf{a}_{n,l}$ equals to one, it indicates that UE $n$ transmits $\mathbf{s}_{k}$ in preamble phase $l$. To illustrate the preamble selection matrix and preamble selection vector, an example is depicted in Fig. \ref{fig13} with $K=4$, $L=2$, $N=3$. It is seen that UE $1$ sends $\mathbf{s}_{1}$ in preamble phase $1$, thus $\mathbf{a}_{1,1} = [1,0,0,0]$. We also see that UE $2$ sends $\mathbf{s}_{1}$ in preamble phase $1$ and $\mathbf{s}_{1}$ in preamble phase $2$, which corresponds to $\mathbf{a}_{2} = [1,0,0,0,1,0,0,0]$.

Based on $\mathbf{B}$ obtained in (\ref{eq4}), UE detection, i.e., super preamble detection, is carried out to obtain the estimation of the preamble selection matrix. The output of the UE detection is denoted as $\hat{\mathbf{A}}=[\hat{\mathbf{a}}_{1}^{T}, \hat{\mathbf{a}}_{2}^{T}, . . . , \hat{\mathbf{a}}_{\hat{N}}^{T}]^{T}\in\mathbb{C}^{\hat{N} \times KL}$, where $\hat{\mathbf{a}}_{n}\in\mathbb{C}^{1\times KL}$ ($n=1,2,...,\hat{N}$) is the preamble selection vector corresponds to the $n$-th UE detected by the BS and $\hat{N}$ is the number of  detected UEs. In general, $\hat{\mathbf{A}}$ is not exactly the same as $\mathbf{A}$, instead it should consist of some of the row vectors of $\mathbf{A}$ and few false preamble selection vectors. After the UE detection, channel estimation is implemented with the Moore-Penrose inverse of $\hat{\mathbf{A}}$, which is given by
\begin{align}\label{eq5}
\hat{\mathbf{H}} =\mathbf{B}\hat{\mathbf{A}}^{+},
\end{align}
where $\hat{\mathbf{H}} = [\hat{\mathbf{h}}_{1},..., \hat{\mathbf{h}}_{\hat{N}}]\in\mathbb{C}^{M \times \hat{N}}$ is the estimated channel response matrix.
Three situations could happen to a detected UE, the $n$-th UE for instance, after channel estimation: 1) the $n$-th detected UE is an actual transmitting UE and ${\mathbf{a}}_{n}$ is not a linear combination of the other row vectors of $\mathbf{A}$, then $\hat{\mathbf{h}}_{n}$ is a valid channel estimation; 2) the $n$-th detected UE is an actual transmitting UE but ${\mathbf{a}}_{n}$ is a linear combination of the other row vectors of $\mathbf{A}$, in this case $\hat{\mathbf{h}}_{n}$ could be erroneous; 3) the $n$-th detected UE is a false UE, the Euclidean norm of $\hat{\mathbf{h}}_{n}$ would be small in general and it thus could be identified and eliminated. We define the RA UE that its preamble selection vector is not a linear combination of the preamble selection vectors of the other $N-1$ RA UEs as a \emph{solvable user}. Then, it is plain that if a solvable user is detected, its channel estimation is valid.

\begin{figure}[!t]
\centering
\includegraphics[width=2.7in]{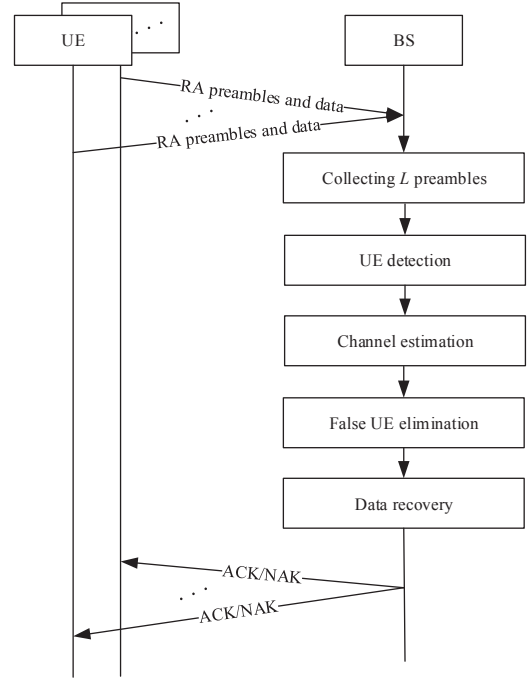}
\caption {The multiple-preamble grant-free RA procedure.} \label{fig12}
\end{figure}

The process at the BS of the proposed RA with super preamble is summarized in Fig. \ref{fig12}. After collecting the $L$ preambles, UE detection is carried out to obtain the estimation of the preamble selection matrix. Then, channel estimation is implemented according to (\ref{eq5}). Evaluating the Euclidean norm of each estimated channel response, false UEs and their channel estimation could be identified and eliminated. With the valid channel estimation, data recovery of detected solvable RA UEs would be successful \cite{Ding2018}.

In summary, the success of UE detection and channel estimation for a RA UE in the proposed approach mainly depend on two conditions: 1) it is a solvable user; 2) its super preamble is detected, i.e., its preamble selection vector is contained in $\hat{\mathbf{A}}$. Therefore, \emph{single user success rate} is defined as the probability that one RA UE is solvable and its super preamble is detected. We are also interested in the \emph{all user success rate}, which is defined as the probability that the preamble selection matrix is full row rank and all the super preambles are detected. Since the goal is to achieve high success rate with the proposed approach, there are two issues remained to be answered,
\begin{itemize}
  \item Do multiple preambles increase the solvable rate effectively? The answer is yes and the analysis will be presented in Section III.

  \item Designing a reliable UE detection method, which will be proposed in Section IV.
\end{itemize}

\section{Solvable Rate Analysis}
In this section, the solvable rates of both single user and all users are analyzed. In the analysis, $N$ UEs simultaneously perform RA and each UE transmits a super preamble consisting of $L$ preambles. In each preamble phase, each RA UE randomly selects a preamble sequence from a pool of $K$ orthogonal preamble sequences.

\subsection{Single User Solvable Rate}
\emph{Single user solvable rate} is defined as the probability that the preamble selection vector of one RA UE is not a linear combination of the preamble selection vectors of other simultaneous RA UEs and denoted by $P^{'}_{\rm{solvable}}(K,L,N)$. When $N<4$, we derive the exact expression of $P^{'}_{\rm{solvable}}(K,L,N)$. When $N\geq4$, we derive an upper bound and a lower bound for $P^{'}_{\rm{solvable}}(K,L,N)$. We take UE $N$ as the target UE for the analysis of $P^{'}_{\rm{solvable}}(K,L,N)$.


When $N=1$, $P^{'}_{\rm{solvable}}(K,L,1)=1$.

When $N=2$, if and only if $\mathbf{a}_{2}$ is different from $\mathbf{a}_{1}$, UE $2$ is solvable. Thus, $P^{'}_{\rm{solvable}}(K,L,2)= 1-\frac{1}{K^{L}}$, where $\frac{1}{K^{L}}$ is the probability that UE $2$ chooses a specific preamble selection vector, i.e., the vector equal to $\mathbf{a}_{1}$.

When $N=3$, if and only if $\mathbf{a}_{3}$ is different from both $\mathbf{a}_{1}$ and $\mathbf{a}_{2}$, UE $3$ is solvable. The reason is explained as follows. If UE $3$ is a linear combination of $\mathbf{a}_{2}$ and $\mathbf{a}_{1}$, i.e.,
\begin{align}\label{eq132}
\mathbf{a}_{3}=q_{1}\mathbf{a}_{1}+q_{2}\mathbf{a}_{2},
\end{align}
where $q_{1}$ and $q_{2}$ are not all zeros, we have that
\begin{align}\label{eq13}
\mathbf{a}_{3,l}=q_{1}\mathbf{a}_{1,l}+q_{2}\mathbf{a}_{2,l},~\mathrm{for}~l = 1,2,...,L.
\end{align}
As only one element of $\mathbf{a}_{i,l}^T$ equals to one and the rest elements are all zeros, there are three possibilities for $q_{1}$ and $q_{2}$: 1) $q_{1}=1$ and $q_{2}=0$, i.e., $\mathbf{a}_{3,l}=\mathbf{a}_{1,l}$ for all $l$, thus $\mathbf{a}_{3}=\mathbf{a}_{1}$; 2) $q_{1}=0$ and $q_{2}=1$, i.e., $\mathbf{a}_{3,l}=\mathbf{a}_{2,l}$ for all $l$, thus $\mathbf{a}_{3}=\mathbf{a}_{2}$; 3) $q_{1}\neq0$ and $q_{2}\neq0$, i.e., $\mathbf{a}_{3,l}=\mathbf{a}_{2,l}=\mathbf{a}_{1,l}$ for all $l$, thus $\mathbf{a}_{3}=\mathbf{a}_{2}=\mathbf{a}_{1}$. As a conclusion, if $\mathbf{a}_{3} \neq \mathbf{a}_{1}$ and $\mathbf{a}_{3} \neq \mathbf{a}_{2}$, UE $3$ is solvable. As the probability that $\mathbf{a}_{3}$ is different from $\mathbf{a}_{1}$ (and $\mathbf{a}_{2}$) is $1-\frac{1}{K^{L}}$,
$P^{'}_{\rm{solvable}}(K,L,3)= (1-\frac{1}{K^{L}})^{2}$.

When $N\geq4$, we derive an upper bound and a lower bound as follows.

The probability that $\mathbf{a}_{N}$ is distinct from $\{\mathbf{a}_{1}$, $\mathbf{a}_{2}$,..., $\mathbf{a}_{N-1}\}$, i.e., none of $\{\mathbf{a}_{1}$, $\mathbf{a}_{2}$,..., $\mathbf{a}_{N-1}\}$ is the same as $\mathbf{a}_{N}$, is one upper bound for $P^{'}_{\rm{solvable}}(K,L,N)$ and it is expressed as
\begin{align}\label{eq201}
P^{'}_{U}(K,L,N) = (1-\frac{1}{K^{L}})^{N-1}.
\end{align}
The simulation results in section V will show that $P^{'}_{\rm{U}}(K,L,N)$ is a very good approximation for $P^{'}_{\rm{solvable}}(K,L,N)$.

To derive the lower bound, we give a proposition firstly in the following.
\newtheorem{Prop}{Proposition}
\begin{Prop}
The number of choices of $\mathbf{a}_{N}$, satisfying the condition that $\mathbf{a}_{N}$ is a linear combination of $\mathbf{a}_{1}$, $\mathbf{a}_{2}$,..., $\mathbf{a}_{N-1}$, is no larger than $\big(\lceil\frac{N-1}{2}\rceil\big)^{L}$.
\end{Prop}
\begin{proof}
If $\mathbf{a}_{N}$ is a linear combination of $\mathbf{a}_{1}$, $\mathbf{a}_{2}$,..., $\mathbf{a}_{N-1}$, i.e.,
\begin{align}\label{eq141}
\mathbf{a}_{N}=q_{1}\mathbf{a}_{1}+q_{2}\mathbf{a}_{2}+...+q_{N-1}\mathbf{a}_{N-1},
\end{align}
where $q_{1}$, $q_{2}$,..., $q_{N-1}$ are not all zeros, we have
\begin{align}\label{eq151}
\mathbf{a}_{N,l}=q_{1}\mathbf{a}_{1,l}+q_{2}\mathbf{a}_{2,l}+...+q_{N-1}\mathbf{a}_{N-1,l},
\end{align}
for preamble phase $l = 1,2,...,L$. Let $q_{1}$, $q_{2}$,..., $q_{N^{'}}$ be the nonzero elements of $\{q_{1}$, $q_{2}$,..., $q_{N-1}\}$ in (\ref{eq151}) without of loss of generality, where $N^{'}\leq N-1$. Then, (\ref{eq151}) is rewritten as
\begin{align}\label{eq131}
\mathbf{a}_{N,l}=q_{1}\mathbf{a}_{1,l}+q_{2}\mathbf{a}_{2,l}+...+q_{N^{'}}\mathbf{a}_{N^{'},l}.
\end{align}

Please be noted that there is only one element in $\mathbf{a}_{n,l}$ ($n=1,2,...,N$) equal to one and the rest elements are all zeros. Based on this fact, we define base vectors among $\{\mathbf{a}_{1,l}$, $\mathbf{a}_{2,l}$, ...,$\mathbf{a}_{N^{'},l}\}$ be a group of vectors that each of them differs from the others and each of $\{\mathbf{a}_{1,l}$, $\mathbf{a}_{2,l}$, ...,$\mathbf{a}_{N^{'},l}\}$ is equivalent to one of them. The number of base vectors among $\{\mathbf{a}_{1,l}$, $\mathbf{a}_{2,l}$, ...,$\mathbf{a}_{N^{'},l}\}$ is denoted as $N_{\rm{b}}$. We also define unique vectors among $\{\mathbf{a}_{1,l}$, $\mathbf{a}_{2,l}$, ...,$\mathbf{a}_{N^{'},l}\}$ be a group of vectors that each of them differs from the other $N^{'}-1$ vectors of $\{\mathbf{a}_{1,l}$, $\mathbf{a}_{2,l}$, ...,$\mathbf{a}_{N^{'},l}\}$. The number of unique vectors among $\{\mathbf{a}_{1,l}$, $\mathbf{a}_{2,l}$, ...,$\mathbf{a}_{N^{'},l}\}$ is denoted as $N_{\rm{u}}$. Then, according to these definitions, it is not difficult to see that
\begin{align}\label{eq601}
 N_{\rm{b}} \leq \big\lfloor (N^{'}-N_{\rm{u}})/2 \big\rfloor + N_{\rm{u}}.
\end{align}

It is noted that $N_{\rm{u}}$ must be no larger than one in order to satisfy (\ref{eq131}), which is proved in the followings. Assuming that the unique vectors among $\{\mathbf{a}_{1,l}$, $\mathbf{a}_{2,l}$, ...,$\mathbf{a}_{N^{'},l}\}$ is $\{\mathbf{a}_{1,l}$, $\mathbf{a}_{2,l}$,..., $\mathbf{a}_{N_{\rm{u}},l}\}$ without of loss of generality, i.e., each $\mathbf{a}_{n,l}$ ($n=1,2,...,N_{\rm{u}}$) is different from the other $N^{'}-1$ vectors of $\{\mathbf{a}_{1,l}$, $\mathbf{a}_{2,l}$, ...,$\mathbf{a}_{N^{'},l}\}$. Then, (\ref{eq131}) is rewritten as
\begin{align}\label{eq152}
\mathbf{a}_{N,l}=q_{1}\mathbf{a}_{1,l}+q_{2}\mathbf{a}_{2,l}+...+q_{N_{\rm{u}}}\mathbf{a}_{N_{\rm{u}},l}+\mathbf{g},
\end{align}
where
\begin{align}\label{eq15}
\mathbf{g}=q_{N_{\rm{u}}+1}\mathbf{a}_{N_{\rm{u}}+1,l}+...+q_{N^{'}}\mathbf{a}_{N^{'},l}.
\end{align}
Apparently, there is no overlap between the nonzero elements of $\mathbf{g}$ and the nonzero elements of $q_{1}\mathbf{a}_{1,l}+q_{2}\mathbf{a}_{2,l}+...+q_{N_{\rm{u}}}\mathbf{a}_{N_{\rm{u}},l}$, since each of $\{\mathbf{a}_{1,l}$, $\mathbf{a}_{2,l}$,..., $\mathbf{a}_{N_{\rm{u}},l}\}$ is unique. As $q_{1}$, $q_{2}$,..., $q_{N_{\rm{u}}}$ are all not zeros, the right of (\ref{eq152}) contains at least $N_{\rm{u}}$ nonzero elements. As there is only one nonzero element in $\mathbf{a}_{N,l}$, $N_{\rm{u}}$ must be no larger than one in order to hold (\ref{eq152}).

Due to (\ref{eq601}) and $N_{\rm{u}} \leq 1$, we have $N_{\rm{b}} \leq \lceil N^{'}/2 \rceil$, i.e., the number of base vectors among $\{\mathbf{a}_{1,l}$, $\mathbf{a}_{2,l}$, ...,$\mathbf{a}_{N^{'},l}\}$ must be no larger than $\lceil N^{'}/2 \rceil$.


As $N^{'}\leq N-1$, $N_{\rm{b}}$ is no larger than $\lceil\frac{N-1}{2}\rceil$.

As $\mathbf{a}_{N,l}$ has to be equivalent to one of base vectors among $\{\mathbf{a}_{1,l}$, $\mathbf{a}_{2,l}$, ...,$\mathbf{a}_{N^{'},l}\}$ (otherwise (\ref{eq151}) never holds), the number of choices of $\mathbf{a}_{N,l}$ satisfying (\ref{eq151}) is no larger than $\lceil\frac{N-1}{2}\rceil$.

Considering the super preamble, the number of choices of $\mathbf{a}_{N}$ satisfying (\ref{eq141}) is no larger than $\big(\lceil\frac{N-1}{2}\rceil\big)^{L}$.

We conclude the proof.
\end{proof}

According to the proposition, we have that
\begin{align}\label{eq23}
1- P^{'}_{\rm{solvable}}(K,L,N)\leq \frac{\big(\lceil\frac{N-1}{2}\rceil\big)^{L}}{K^{L}},
\end{align}
then,
\begin{align}\label{eq24}
P^{'}_{\rm{solvable}}(K,L,N)\geq 1-\frac{\big(\lceil\frac{N-1}{2}\rceil\big)^{L}}{K^{L}}.
\end{align}
As a consequence, we obtain the lower bound for $P^{'}_{\rm{solvable}}(K,L,N)$ as
\begin{align}\label{eq231}
P^{'}_{L}(K,L,N) = 1-\frac{\big(\lceil\frac{N-1}{2}\rceil\big)^{L}}{K^{L}}.
\end{align}
Please be noted that $P^{'}_{L}(K,L,N)$ is a very loose lower bound for $P^{'}_{\rm{solvable}}(K,L,N)$ as we do not consider the constraint that $\mathbf{a}_{N,l}$ must satisfy (\ref{eq151}) with a same set of $\{q_{1}$, $q_{2}$,..., $q_{N-1}\}$ for all $l$ in the derivation. If we take into consideration of this constraint, the number of choices of $\mathbf{a}_{N}$ satisfying (\ref{eq141}) should be much less than $\big(\lceil\frac{N-1}{2}\rceil\big)^{L}$.

\newtheorem{remark}{Remark}
\begin{remark}
It is clear from {\rm{(\ref{eq231})}} that $\frac{\big(\lceil\frac{N-1}{2}\rceil\big)^{L}}{K^{L}}$ approaches to zero rapidly as $L$ increases under the condition that $K\gg\lceil\frac{N-1}{2}\rceil$, i.e., the lower bound of $P^{'}_{\rm{solvable}}(K,L,N)$ approaches to one effectively as $L$ increases. Therefore, it is concluded that using multiple preambles is very effective in increasing $P^{'}_{\rm{solvable}}(K,L,N)$.
\end{remark}

\subsection{All User Solvable Rate}
\emph{All user solvable rate} is defined as the probability that $\mathbf{A}$ is full row rank, and denoted by $P_{\rm{solvable}}(K,L,N)$.
When $N<4$, we derive the exact expression of $P_{\rm{solvable}}(K,L,N)$. When $N\geq4$, we derive an upper bound and a lower bound for $P_{\rm{solvable}}(K,L,N)$.

When $N=1$, $P_{\rm{solvable}}(K,L,1)=1$.

When $N=2$, if and only if $\mathbf{a}_{1} \neq \mathbf{a}_{2}$, UE$1$ and UE $2$ are both solvable. Thus, $P_{\rm{solvable}}(K,L,2) = 1-\frac{1}{K^{L}}$.

When $N=3$, if and only if $\mathbf{a}_{1}$, $\mathbf{a}_{2}$, $\mathbf{a}_{3}$ are different from each other, UE $1$, UE $2$ and UE $3$ are all solvable. Thus, $P_{\rm{solvable}}(K,L,3) = (1-\frac{1}{K^{L}})(1-\frac{2}{K^{L}})$.

When $N\geq4$, we derive an upper bound and a lower bound as follows.

The probability that $\mathbf{a}_{1}$, $\mathbf{a}_{2}$,..., $\mathbf{a}_{N}$ are different from each other is one upper bound for $P_{\rm{solvable}}(K,L,N)$ and it is given as
\begin{align}\label{eq13}
&P_{\rm{U}}(K,L,N) = \nonumber\\
&\frac{(K^{L}-1)(K^{L}-2)(K^{L}-3)...\big(K^{L}-(N-1)\big)}{K^{(N-1)L}}.
\end{align}
The simulation results in section V will show that $P_{\rm{U}}(K,L,N)$ is a very good approximation for $P_{\rm{solvable}}(K,L,N)$.

If the matrix $[\mathbf{a}_{1}, \mathbf{a}_{2},..., \mathbf{a}_{N-1}]$ is full row rank and $\mathbf{a}_{N}$ is not a linear combination of $\mathbf{a}_{1}$, $\mathbf{a}_{2}$,..., $\mathbf{a}_{N-1}$, $\mathbf{A}$ is full row rank. Thus, $P_{\rm{solvable}}(K,L,N)$ can be expressed as
\begin{align}\label{eq132}
&P_{\rm{solvable}}(K,L,N)=\nonumber\\
&P_{\rm{solvable}}(K,L,N-1)\big( 1-P^{''}(K,L,N) \big),
\end{align}
where $P^{''}(K,L,N)$ is the probability that $\mathbf{a}_{N}$ is a linear combination of $\mathbf{a}_{1}$, $\mathbf{a}_{2}$,..., $\mathbf{a}_{N-1}$ under the condition that the matrix $[\mathbf{a}_{1}, \mathbf{a}_{2},..., \mathbf{a}_{N-1}]$ is full row rank. Proposition 1 also applies to estimation of $P^{''}(K,L,N)$, then we have that
\begin{align}\label{eq211}
P^{''}(K,L,N)\leq \frac{\big( \lceil\frac{N-1}{2}\rceil \big)^{L}}{K^{L}}.
\end{align}
Combining (\ref{eq132}) and (\ref{eq211}), we have that
\begin{align}\label{eq182}
&P_{\rm{solvable}}(K,L,N)\geq \nonumber\\
&P_{\rm{solvable}}(K,L,N-1)\bigg(1-\frac{\big( \lceil\frac{N-1}{2}\rceil \big)^{L}}{K^{L}}\bigg).
\end{align}
Since the exact expression of $P_{\rm{solvable}}(K,L,3)$ is available, (\ref{eq182}) is further derived as
\begin{align}\label{eq18}
&P_{\rm{solvable}}(K,L,N)\geq \nonumber\\
&P_{\rm{solvable}}(K,L,3)\bigg(1 - \frac{\big( \lceil\frac{4-1}{2}\rceil \big)^{L}}{K^{L}}\bigg)...\bigg(1-\frac{\big( \lceil\frac{N-1}{2}\rceil \big)^{L}}{K^{L}}\bigg).
\end{align}
Thus, we obtain the lower bound for $P_{\rm{solvable}}(K,L,N)$ as
\begin{align}\label{eq191}
&P_{L}(K,L,N)= \nonumber\\
&P_{\rm{solvable}}(K,L,3)\bigg(1 - \frac{\big( \lceil\frac{4-1}{2}\rceil \big)^{L}}{K^{L}}\bigg)...\bigg(1-\frac{\big( \lceil\frac{N-1}{2}\rceil \big)^{L}}{K^{L}}\bigg).
\end{align}

\begin{remark}
Similar to the case of single user solvable rate, it is concluded that using multiple preambles is very effective in increasing $P_{\rm{solvable}}(K,L,N)$.
\end{remark}

\section{UE Detection with the Support of Massive MIMO}
In each preamble phase, the BS is able to detect the preambles by performing the following operation: if the Euclidean norm of the $k$th column vector of $\mathbf{B}_{l}$ in (\ref{eq2}) is higher than a predefined threshold, the BS would determine that at least one UE has transmitted the preamble sequence $\mathbf{s}_{k}$ in preamble phase $l$. However, with single antenna or small number of antennas, the BS is unable to determine which two preamble sequences that respectively belong to two different preamble phases are transmitted by a same RA UE. In other words, it is difficult for the BS to detect the super preamble of a RA UE, i.e., obtaining the preamble selection vector of the RA UE.

It is very different in the case of massive MIMO, where the channels of any two RA UEs are quasi orthogonal, i.e., have close-to-zero spatial correlation. On the other hand, the preamble signals that transmitted by one RA UE in two different preamble phases, may be different sequences, should be strongly correlated in space, in the case of massive MIMO. By exploiting this quasi-orthogonality characteristic, the BS is able to determine which two preamble sequences that respectively belong to two different preamble phases are transmitted by one RA UE.

For preamble phase $l$ and $l^{'}$, where $l<l^{'}$ ($l, l^{'}\in\{1,2,...,L\}$), we correlate $\mathbf{B}_{l}$ with $\mathbf{B}_{l^{'}}$, where $\mathbf{B}_{l}$ and $\mathbf{B}_{l^{'}}$ could be obtained according to (\ref{eq2}). The correlation result of $\mathbf{B}_{l}$ and $\mathbf{B}_{l^{'}}$, denoted by $\mathbf{C}_{l,l^{'}}\in\mathbb{C}^{K \times K}$, is given by
\begin{align}\label{eq29}
\mathbf{C}_{l,l^{'}} = \mathbf{B}_{l}^{H}\mathbf{B}_{l^{'}}.
\end{align}
Using (\ref{eq3}), we have that
\begin{align}\label{eq25}
\mathbf{C}_{l,l^{'}} = \mathbf{A}^{H}_{l}{\mathbf{H}}^{H}{\mathbf{H}}\mathbf{A}_{l^{'}}+\mathbf{A}^{H}_{l}{\mathbf{H}}^{H}\mathbf{W}_{l^{'}}+\nonumber\\
\mathbf{W}^{H}_{l}\mathbf{H}\mathbf{A}_{l^{'}}+\mathbf{W}^{H}_{l}\mathbf{W}_{l^{'}}.
\end{align}
Considering that $\mathbf{W}_{l}$ and $\mathbf{W}_{l^{'}}$ are independent from $\mathbf{H}$, $\mathbf{A}^{H}_{l}{\mathbf{H}}^{H}\mathbf{W}_{l^{'}}$ and $\mathbf{W}^{H}_{l}\mathbf{H}\mathbf{A}_{l^{'}}$ should be much less than $\mathbf{A}^{H}_{l}{\mathbf{H}}^{H}{\mathbf{H}}\mathbf{A}_{l^{'}}$ in average power as the gain of massive MIMO is high, thus $\mathbf{A}^{H}_{l}{\mathbf{H}}^{H}\mathbf{W}_{l^{'}}$ and $\mathbf{W}^{H}_{l}\mathbf{H}\mathbf{A}_{l^{'}}$ could be ignored. Also considering that $\mathbf{W}_{l}$ is independent from $\mathbf{W}_{l^{'}}$, $\mathbf{W}^{H}_{l}\mathbf{W}_{l^{'}}$ should also be much less than $\mathbf{A}^{H}_{l}{\mathbf{H}}^{H}{\mathbf{H}}\mathbf{A}_{l^{'}}$ in average power as the gain of massive MIMO is high. Thus, $\mathbf{W}^{H}_{l}\mathbf{W}_{l^{'}}$ could also be ignored. Then, we have the approximation as
\begin{align}\label{eq255}
\mathbf{C}_{l,l^{'}} \approx \mathbf{A}^{H}_{l}{\mathbf{H}}^{H}{\mathbf{H}}\mathbf{A}_{l^{'}}.
\end{align}
In massive MIMO system, the channels of any two uplink RA UEs could be assumed quasi orthogonal, i.e., ${\mathbf{H}}^{H}{\mathbf{H}}\approx\mathbf{I}$. Then, (\ref{eq255}) is rewritten as
\begin{align}\label{eq26}
\mathbf{C}_{l,l^{'}} \approx \mathbf{A}_{l}^{H}\mathbf{A}_{l^{'}},
\end{align}
where the $\mu$-th row and $\nu$-th column element of $\mathbf{C}_{l,l^{'}}$ is
\begin{align}\label{eq271}
(\mathbf{C}_{l,l^{'}})_{\mu,\nu} \approx \big((\mathbf{A}_{l})_{-,\mu}\big)^{H}(\mathbf{A}_{l^{'}})_{-,\nu}.
\end{align}
If UE $n$ transmits the $\mu$-th preamble sequence at preamble phase $l$ and the $\nu$-th preamble sequence at preamble phase $l^{'}$, $(\mathbf{A}_{l})_{-,\mu}$ and $(\mathbf{A}_{l^{'}})_{-,\nu}$ both have an element with value of one in the $n$-th row. Then, it is a high probability that $(\mathbf{C}_{l,l^{'}})_{\mu,\nu}$ is large, here we consider  $(\mathbf{C}_{l,l^{'}})_{\mu,\nu}$ as large if $|(\mathbf{C}_{l,l^{'}})_{\mu,\nu}|$ is larger than a predefined threshold. We will discuss how to set this threshold in the following paragraphes. Based on this observation, we could use $(\mathbf{C}_{l,l^{'}})_{\mu,\nu}$, $1\leq\mu,\nu \leq K$, to determine which two preamble sequences are transmitted by UE $n$ respectively in preamble phase $l$ and $l^{'}$.

\begin{figure}[!t]
\centering
\includegraphics[width=2.2 in]{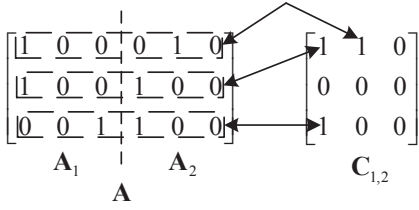}
\caption {$\mathbf{A}$ and $\mathbf{C}_{12}$ with $K=3$, $L=2$, $N=3$.} \label{fig14}
\end{figure}

To further illustrate (\ref{eq271}), an example is depicted in Fig. \ref{fig14} with $L=2$, $K=3$, $N=3$. It is observed that UE 1 transmits the 1-st preamble sequence at preamble phase 1 and the 2-nd preamble sequence at preamble phase 2, hence $(\mathbf{C}_{1,2})_{1,2}$ is large, where $(\mathbf{C}_{1,2})_{1,2}=\big((\mathbf{A}_{1})_{-,1}\big)^{H}(\mathbf{A}_{2})_{-,2}$. We also observe that UE 2 transmits the 1-st preamble sequence at preamble phase 1 and the 1-st preamble sequence at preamble phase 2, hence $(\mathbf{C}_{1,2})_{1,1}$ is large, where $(\mathbf{C}_{1,2})_{1,1}=\big((\mathbf{A}_{1})_{-,1}\big)^{H}(\mathbf{A}_{2})_{-,1}$. For UE 3, we have the similar observation as UE 1 and UE 2. As each large $\mathbf{C}_{l,l^{'}}$ corresponds to a RA UE, we could use $\mathbf{C}_{l,l^{'}}$ to determine which two preamble sequences from two different preamble phases are transmitted by one RA UE, i.e., we could use $\mathbf{C}_{1,2}$ to acquire the preamble selection vector of each RA UE in Fig. \ref{fig14}. For example, $(\mathbf{C}_{1,2})_{1,1}=1$ indicates that one UE transmits the $1$st preamble sequence in preamble phase $1$ and the $1$st preamble sequence at preamble phase $2$. Therefore, we could acquire a preamble selection vector as $[1,0,0,1,0,0]$. Similarly, we could also acquire a preamble selection vector as $[1,0,0,0,1,0]$ corresponding to $(\mathbf{C}_{1,2})_{1,2}$ and a preamble selection vector as $[0,0,1,1,0,0]$ corresponding to $(\mathbf{C}_{1,2})_{3,1}$. With these three preamble selection vectors, we could form $\hat{\mathbf{A}}$ as
\begin{align}\label{eq272}
\hat{\mathbf{A}} =
\left[
  \begin{array}{cccccc}
    1 & 0 & 0 & 1 & 0 & 0 \\
    1 & 0 & 0 & 0 & 1 & 0 \\
    0 & 0 & 1 & 1 & 0 & 0
  \end{array}
\right].
\end{align}
It is obvious that $\hat{\mathbf{A}}$ could be a row switching transformation of $\mathbf{A}$.

For any $L\geq2$, let $\{\theta_{1}, \theta_{2},..., \theta_{L}\}$ denote the indexes of the preamble sequences that one UE transmits in the $L$ preamble phases, where $\theta_{l}\in \{1,2,...,K\}$ for $l=1,2,...,L$. Then, it is a high probability that
\begin{align}\label{eq281}
|(\mathbf{C}_{l,l^{'}})_{\theta_{l},\theta_{l^{'}}}|>\mathrm{TH},~\mathrm{for~all}~l~\mathrm{and}~l^{'},
\end{align}
where $l<l^{'}$, $l, l^{'}\in\{1,2,...,L\}$ and TH represent the threshold. On the other hand, if there is no UE transmit the $\theta_{l}$-th preamble sequence at the $l$-th preamble phase and the $\theta_{l^{'}}$-th preamble sequence at the $l^{'}$-th preamble phase, $|(\mathbf{C}_{l,l^{'}})_{\theta_{l},\theta_{l^{'}}}|$ approximately equals to zero due to the spatial quasi-orthogonality between channels of different UEs. Thus, it is not difficult to find a proper threshold (TH) to separate these two cases.

Based on (\ref{eq281}), the UE detection at the BS is simply the exhaustive search among all the $K^L$ choices of $\{\theta_{1}, \theta_{2},..., \theta_{L}\}$ to pick out the choices that satisfy (\ref{eq281}), each of which corresponds to the preamble selection vector of a possible RA UE. With the obtained preamble selection vectors, $\hat{\mathbf{A}}$ is formed, which is an estimation of the preamble selection matrix. The details of the proposed UE detection algorithm are presented as Algorithm 1.

\renewcommand{\algorithmicrequire}{\textbf{Input:}}  
\renewcommand{\algorithmicensure}{\textbf{Output:}}  
\begin{algorithm}[htb]
\label{algorithm_1} \caption{UE detection}
\begin{algorithmic}[1]

\REQUIRE~~
    $K$, ~$L$, ~$\mathbf{B}$, ~$\mathrm{TH}$;
    \vspace{5pt}

\STATE $n=1$

\FOR{$\theta_{1}=1$ to $K$;~$\theta_{2}=1$ to $K$;~\ldots~;~$\theta_{L}=1$ to $K$ }

        \IF { $\{\theta_{1},\theta_{2},...,\theta_{L}\}$ satisfies (\ref{eq281})}

           \STATE {Add a new row to $\hat{\mathbf{A}}$ and initialize it to all zeros:
           $(\hat{\mathbf{A}})_{n,-} = \mathbf{0}\in\mathbb{C}^{1\times KL}$ }
           \FOR{$l=1$ to $L$ }

           \STATE { $(\hat{\mathbf{A}})_{n,\theta_{l}+(l-1)K}=1$ }

           \ENDFOR

              \STATE {$n=n+1$}

        \ENDIF

\ENDFOR

\vspace{5pt}
\ENSURE~~$\hat{\mathbf{A}}$
\end{algorithmic}
\end{algorithm}

A key issue of the proposed algorithm is how to set TH. If TH is set too high, the preamble selection vectors of some RA UEs may not be contained in $\hat{\mathbf{A}}$, which results in miss detection. If TH is set too low, $\hat{\mathbf{A}}$ may contain some false preamble selection vectors, which results in false detection. As we mentioned before in Section II, the false preamble selection vectors could be identified and eliminated, as the column of $\hat{\mathbf{H}}$ in (\ref{eq5}) that corresponds to any false preamble selection vector has a Euclidean norm close to zero. Therefore, a lower TH is preferred in the proposed algorithm to guarantee low miss rate.

Finally, after we obtain the estimation of the preamble selection matrix, the channel estimation of the solvable RA UEs can be obtained according to (\ref{eq5}).

\section{Numerical Results}
In this section, numerical results are presented to verify the effectiveness of the proposed multi-preamble approach, in terms of the solvable rate, the success rate and the normalized mean square error (NMSE) performance of channel estimation. Single user success rate is defined as the probability that one RA UE is solvable and its super preamble is detected, which is denoted as $P^{'}_{\rm{success}}$. All user success rate is defined as the probability that the preamble selection matrix is full row rank and all the super preambles are detected, which is denoted as $P_{\rm{success}}$. The NMSE of channel estimation is defined as
\begin{align}\label{eq31}
{\rm{NMSE}} = \frac{ {\rm{mean}}\big(\|\hat{\mathbf{h}}_{n} -\mathbf{h}_{n}\|^{2}\big) }{ {\rm{mean}}\big(\|\mathbf{h}_{n}\|^{2}\big)},
\end{align}
where $\hat{\mathbf{h}}_{n}$ is the channel estimation result of the $n$-th successful RA UE and $\mathbf{h}_{n}$ is the actual channel of this UE. In simulations, NMSE results are averaged over $10^{5}$ Monte Carlos trials. The signal to noise ratio (SNR) is defined as the preamble to noise power ratio at each antenna port of the BS. TH of UE detection is set as $0.4$. The simulation parameters are summarized in Table I.
\begin{table}[!t]
\renewcommand{\arraystretch}{1}
\caption{Simulation Parameters}\label{table1} \centering
\begin{tabular}{|>{\centering}m{4.5cm}|>{\centering}m{3cm}|}

\hline
Number of antennas $M$ & 128 \tabularnewline
\hline Number of preamble phases $L$  & 1 $\sim$ 6 \tabularnewline
\hline Number of orthogonal preamble sequences $K$ & 8 $\sim$ 48 \tabularnewline
\hline Number of simultaneous RA UEs $N$ & 1 $\sim$ 20 \tabularnewline
\hline SNR & 0 $\sim$ 20 dB \tabularnewline
\hline TH & 0.4 \tabularnewline
\hline

\end{tabular}
\end{table}

We consider two different massive MIMO channel models in the simulations:
\subsubsection{Independent Rayleigh fading Channel}
Propagation between the $M$ base station antennas and $N$ RA UEs is described by an matrix $\sqrt{1/M}\mathbf{H}\in\mathbb{C}^{M \times N}$ , where the entries of $\mathbf{H}$ are independent $\mathcal{CN}(0,1)$ random variables and the coefficient $\sqrt{1/M}$ normalizes the expected power of the channel response vector to 1, i.e., $  \mathbb{E}\{ \|\mathbf{h}_{n}\|^{2} \}=1$. Here, $\mathcal{CN}(0,1)$ denotes circularly-symmetric complex Gaussian distribution with zero-mean and unit-variance.

\subsubsection{Spatially Correlated Rayleigh Fading Channel}
Spatially correlated Rayleigh fading is a more realistic channel model, which has been widely used in MIMO systems for analysis and simulations \cite{Hoydis2013} \cite{Xu2014}. The channel response between the BS and an arbitrary RA UE is modelled by $\mathbf{h}\in \mathbb{C}^{M}$, which is given by,
\begin{align}\label{eq111}
\mathbf{h}=\dfrac{1}{\sqrt{M}}\mathbf{R}\mathbf{v},
\end{align}
where $\mathbf{h}$ stands for small scale fading vector between UE and BS, $\mathbf{R}\in \mathbb{C}^{M\times Q}$ is antenna correlation matrix, $\mathbf{v}\sim \mathcal{CN}(0,\mathbf{I}_{Q})$ is independent fast-fading channel vector, where $Q$ is the number of independently faded paths.

For a uniform linear array, $\mathbf{R}=[\mathbf{r}(\phi_1),\ldots,\mathbf{r}(\phi_Q)]$ is composed of the
steering vector $\mathbf{r}(\phi_q)$ defined as
\begin{align}\label{eq222}
\mathbf{r}(\phi_q)=\frac{1}{\sqrt{Q}}[1,e^{-\textrm{j}2\pi\omega\cos(\phi_q)},\ldots,e^{-\textrm{j}2\pi\omega(M-1)\cos(\phi_q)}]^{T},
\end{align}
where $\phi_q$ ($q=1,\ldots,Q$) is the angle of arrival (AOA) of the $q$th path, which is uniformly generated within $[\phi_\mathrm{A}-\frac{\phi_\mathrm{S}}{2}, \phi_\mathrm{A}+\frac{\phi_\mathrm{S}}{2}]$.
And $\phi_\mathrm{A}$ and $\phi_\mathrm{S}$ are defined as the azimuth angle of the UE location and the angle spread, respectively. $\omega$ is the antenna spacing in multiples of the wavelength. The parameters of spatially correlated Rayleigh fading channel are given in Table II.

\begin{table}[!t]
\renewcommand{\arraystretch}{1}
\caption{Simulation Parameters of Spatially Correlated Fading Channel}\label{table2} \centering
\begin{tabular}{|>{\centering}m{4.5cm}|>{\centering}m{3cm}|}

\hline
Number of faded paths $Q$  & 50 \tabularnewline
\hline Antenna spacing $\omega$ & $1/2$ \tabularnewline
\hline Angle spread $\phi_\mathrm{S}$ & $40^{\circ}$ \tabularnewline
\hline Azimuth angle $\phi_\mathrm{A}$ & uniform distribution within $(-180^{\circ}, 180^{\circ}]$ \tabularnewline \hline

\end{tabular}
\end{table}

\subsection{Solvable Rate}
\begin{figure}[!t]
\centering
\includegraphics[width=3.2in]{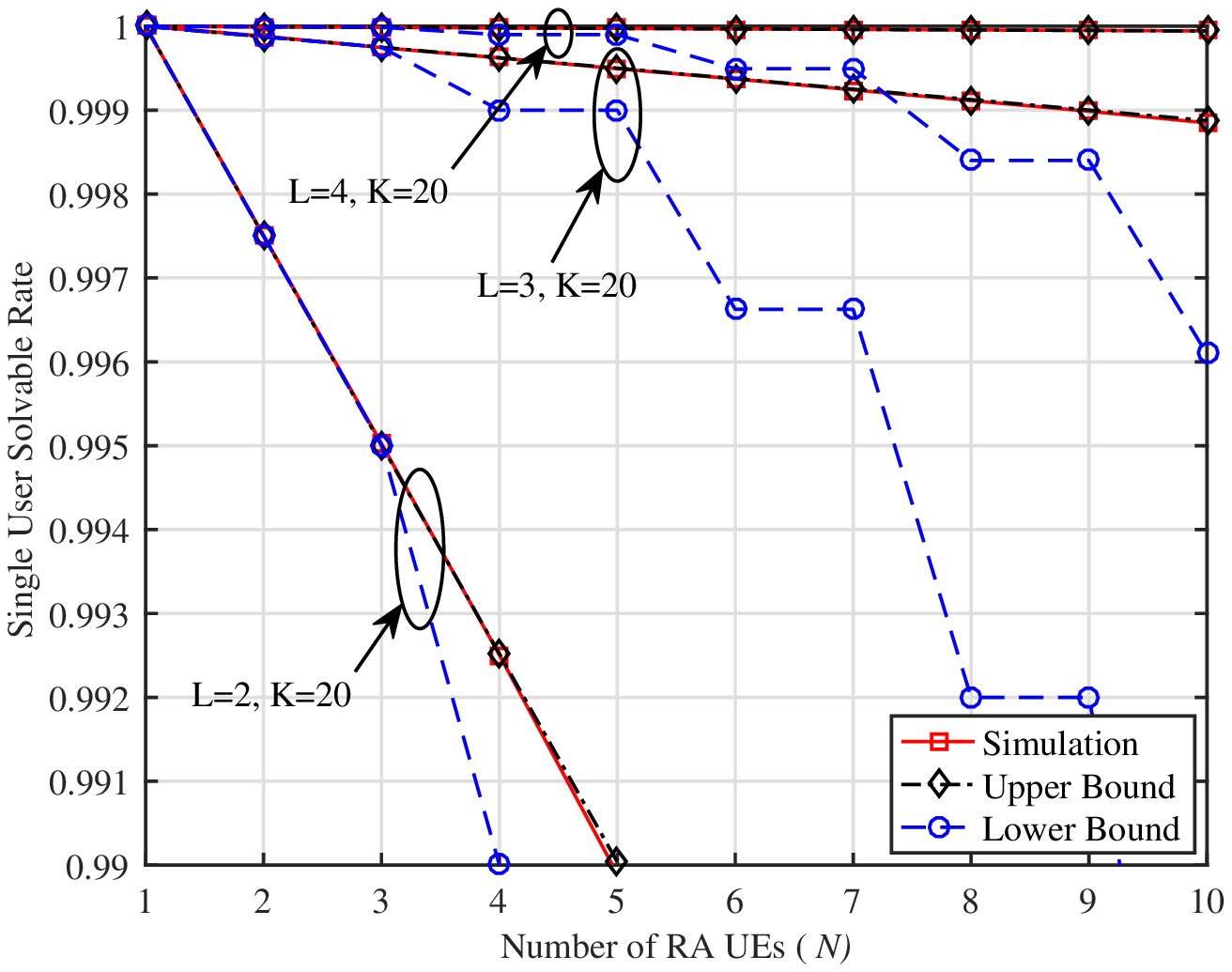}
\caption{$P^{'}_{\rm{solvable}}$ versus $N$ with different sets of $L$ and $K$.} \label{fig2}
\end{figure}

\begin{figure}[!t]
\centering
\includegraphics[width=3.2in]{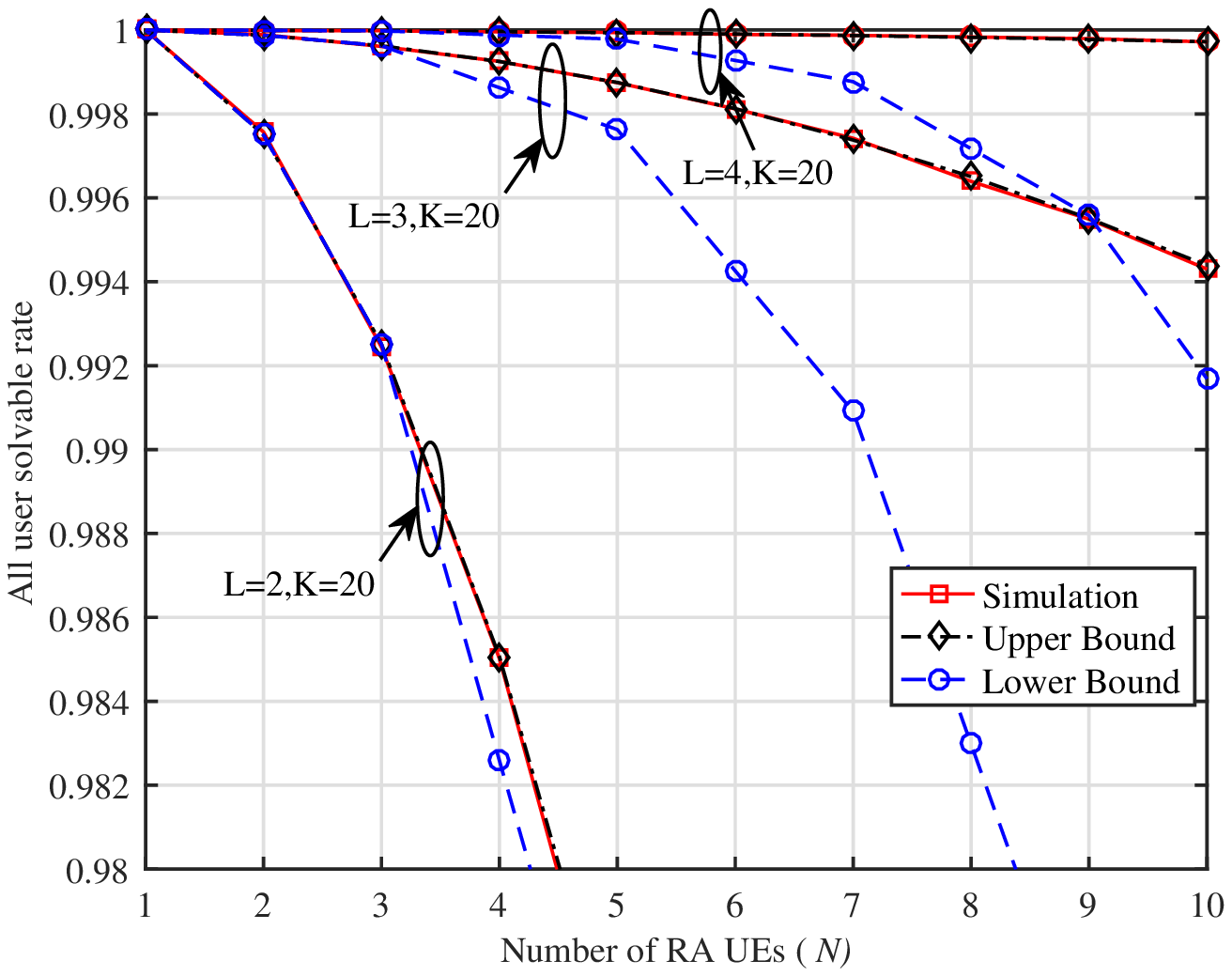}
\caption{$P_{\rm{solvable}}$ versus $N$ with different sets of $L$ and $K$.} \label{fig1}
\end{figure}

In Fig. \ref{fig2}, the simulated single user solvable rate, derived upper bound and lower bound are presented for different sets of $K$ and $L$. The upper bound and lower bound are respectively obtained via (\ref{eq201}) and (\ref{eq231}) in Section III. It is observed that although the lower bound is loose, it approaches to one as $L$ increases. It is also observed that the upper bound is very tight, thus it could be used as a good approximation of the single user solvable rate. From these observations, we could conclude that adding preambles is very effective in increasing the single user solvable rate. We also present the simulated all user solvable rate, derived upper bound and lower bound in Fig. \ref{fig1} and similar observations are obtained.

\subsection{Multiple Preambles versus Single Preamble}
\begin{figure}[!t]
\centering
\includegraphics[width=3.2in]{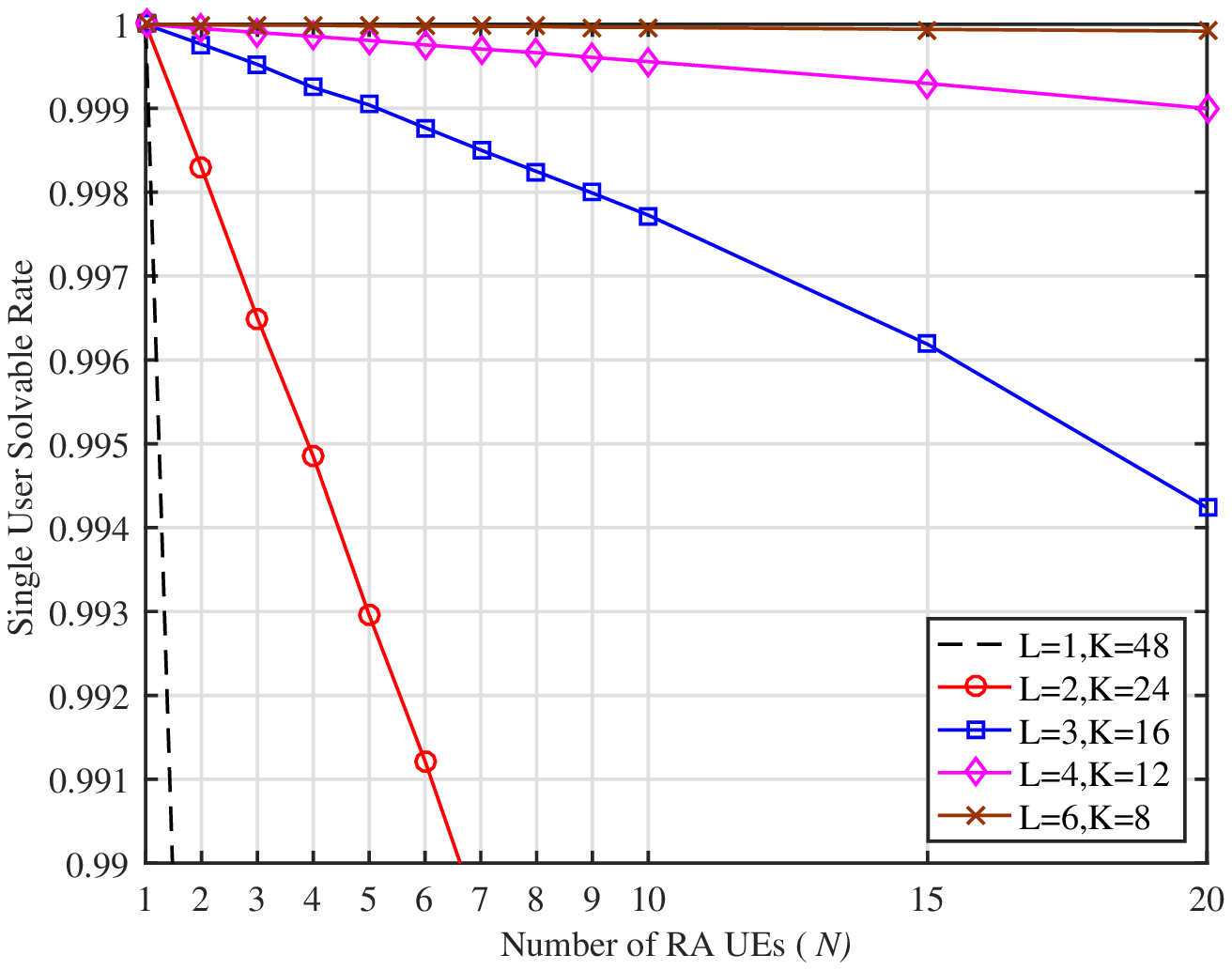}
\caption{$P^{'}_{\rm{solvable}}$ versus $N$ with different $L$ under the constraint that $KL=48$.} \label{fig4}
\end{figure}

\begin{figure}[!t]
\centering
\includegraphics[width=3.2in]{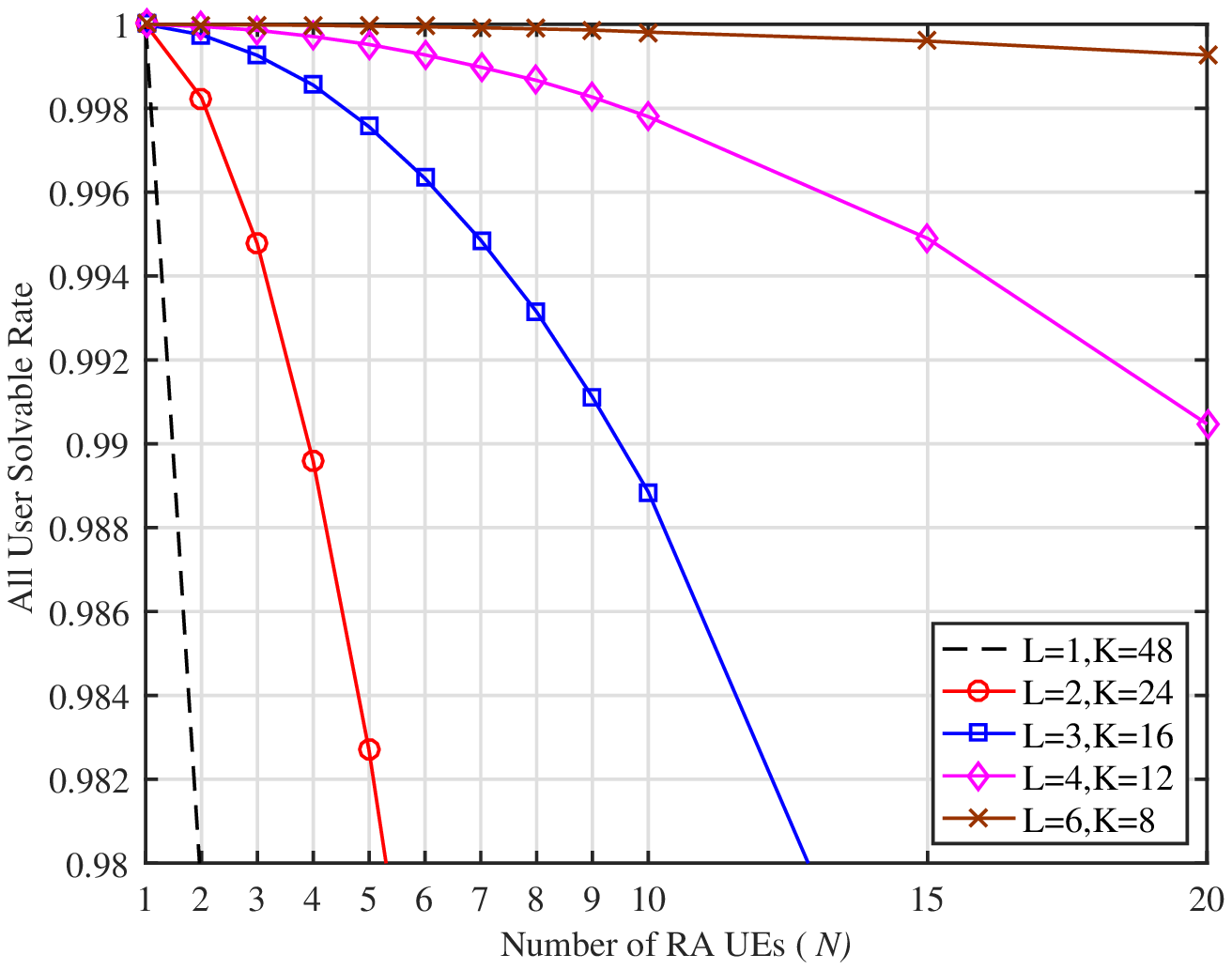}
\caption{$P_{\rm{solvable}}$ versus $N$ with different $L$ under the constraint that $KL=48$.} \label{fig3}
\end{figure}

In Fig. \ref{fig4}, the length of the super preamble remains unchanged as $48$ (i.e., $KL=48$) and the simulation results of $P^{'}_{\rm{solvable}}$ are plotted as a function of $N$ with different $L$. We see that when $L=1$, which respresent the traditional single preamble case, the BS can only serve one RA UE at $P^{'}_{\rm{solvable}}=0.99$. When $L=2$, the number of RA UEs that the BS can simultaneously serve increases to $6$ at $P^{'}_{\rm{solvable}}=0.99$, which is about six times that of $L=1$. Further increasing $L$, the number of RA UEs that the BS can serve at $P^{'}_{\rm{solvable}}=0.99$ keeps rising. It worth noting that the total preamble resources are kept unchanged for different $L$ in the simulations, i.e., the total length of preambles is unchanged ($KL=48$). From these observations, we conclude that using the proposed multi-preamble approach, higher single user solvable rate could be achieved by breaking a single preamble into multiple preambles of shorter length. We also present the simulation results of $P_{\rm{solvable}}$ with constant $KL$ in Fig. \ref{fig3} and similar observations are obtained.

\subsection{Success Rate}
\begin{figure}[!t]
\centering
\includegraphics[width=3.2in]{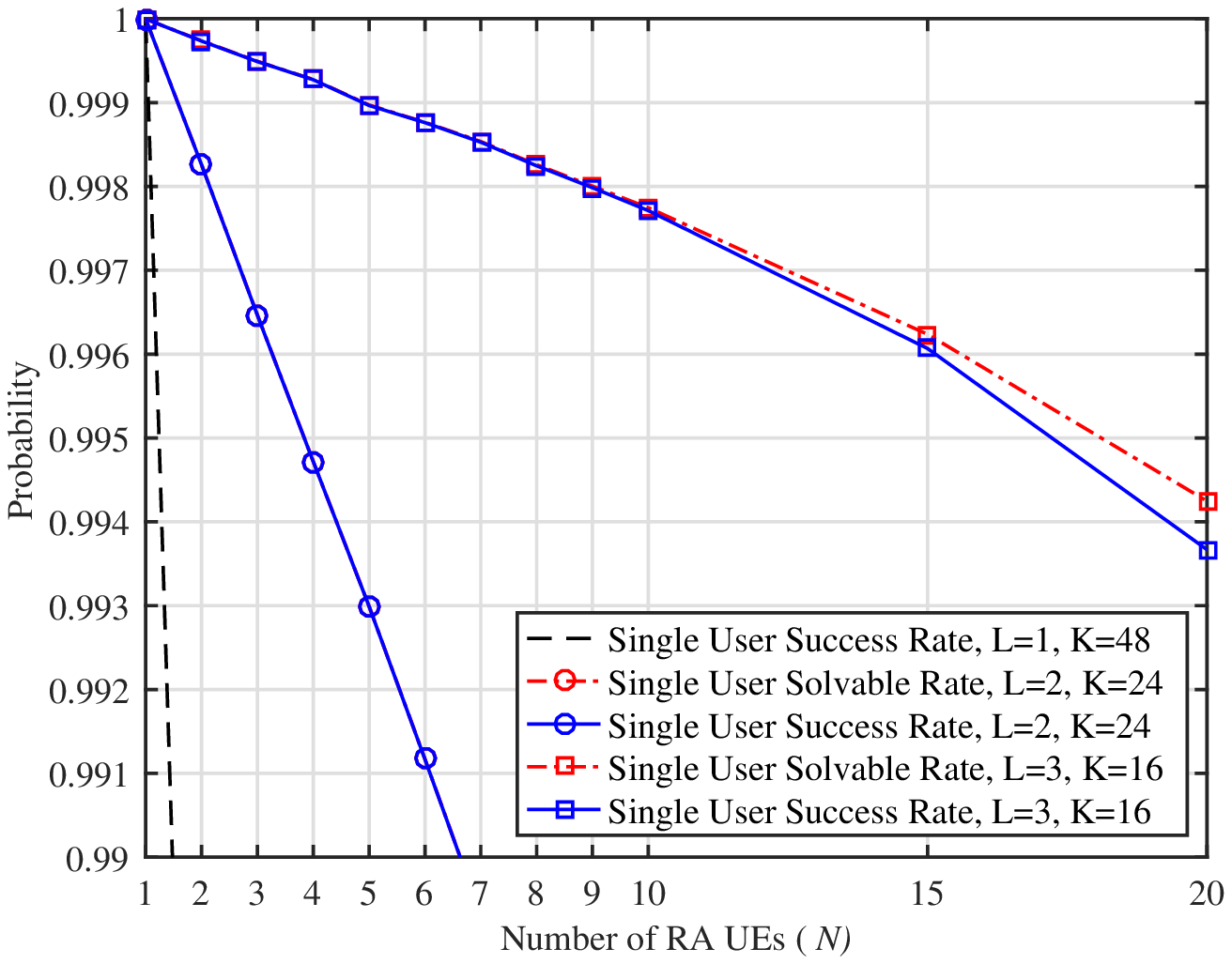}
\caption{$P^{'}_{\rm{success}}$ under independent rayleigh fading channel with $M=128$ and $\rm{SNR}=0$ dB. } \label{fig7}
\end{figure}

\begin{figure}[!t]
\centering
\includegraphics[width=3.2in]{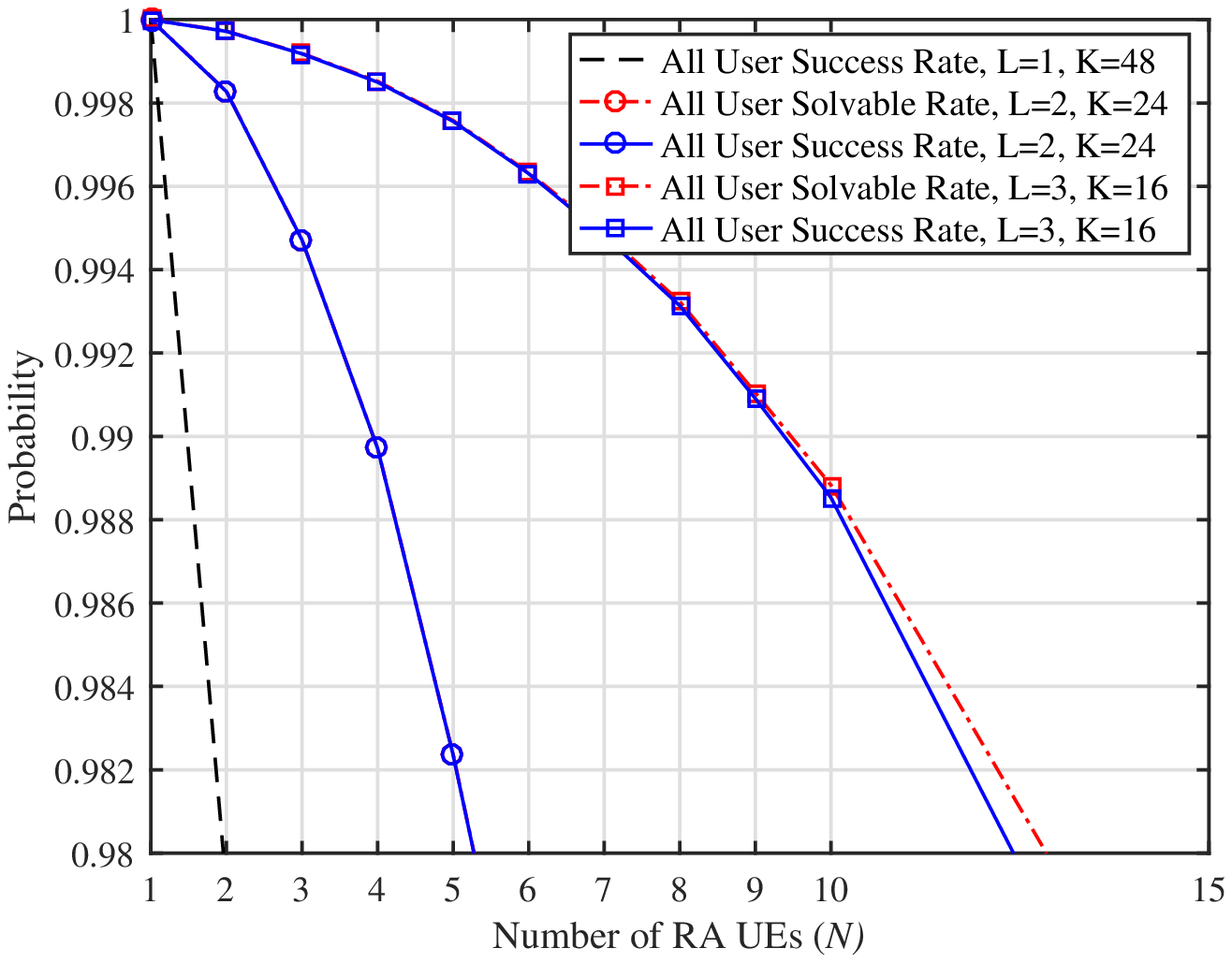}
\caption{$P_{\rm{success}}$ under independent rayleigh fading channel with $M=128$ and $\rm{SNR}=0$ dB.} \label{fig5}
\end{figure}

In Fig. \ref{fig7}, the simulation results of $P^{'}_{\rm{success}}$ are presented as a function of $N$ with different $L$ under independent rayleigh fading channel and $\rm{SNR}=0$ dB. To demonstrate the performance of the proposed UE detection algorithm, Fig. \ref{fig7} also includes the simulated $P^{'}_{\rm{solvable}}$ with $L=2$ and $L=3$. The gap between the curves $P^{'}_{\rm{success}}$ and $P^{'}_{\rm{solvable}}$ is the probability that the BS fails to detect the super preamble of a solvable RA UE. Therefore, a smaller gap indicates a lower miss UE detection rate. It is observed that when $L=2$, the curve of $P^{'}_{\rm{success}}$ coincides with the curve of $P^{'}_{\rm{solvable}}$. When $L=3$, the curve of $P^{'}_{\rm{success}}$ is also very tight to the curve of $P^{'}_{\rm{solvable}}$. These results show that the proposed algorithm has very good performance in UE detection under independent rayleigh fading channel, due to the quasi-orthogonality among channels of RA UEs. Simulations under independent rayleigh fading channel for $P_{\rm{success}}$ and $P_{\rm{solvable}}$ are presented in Fig. \ref{fig5} and similar observations are obtained.

\begin{figure}[!t]
\centering
\includegraphics[width=3.2in]{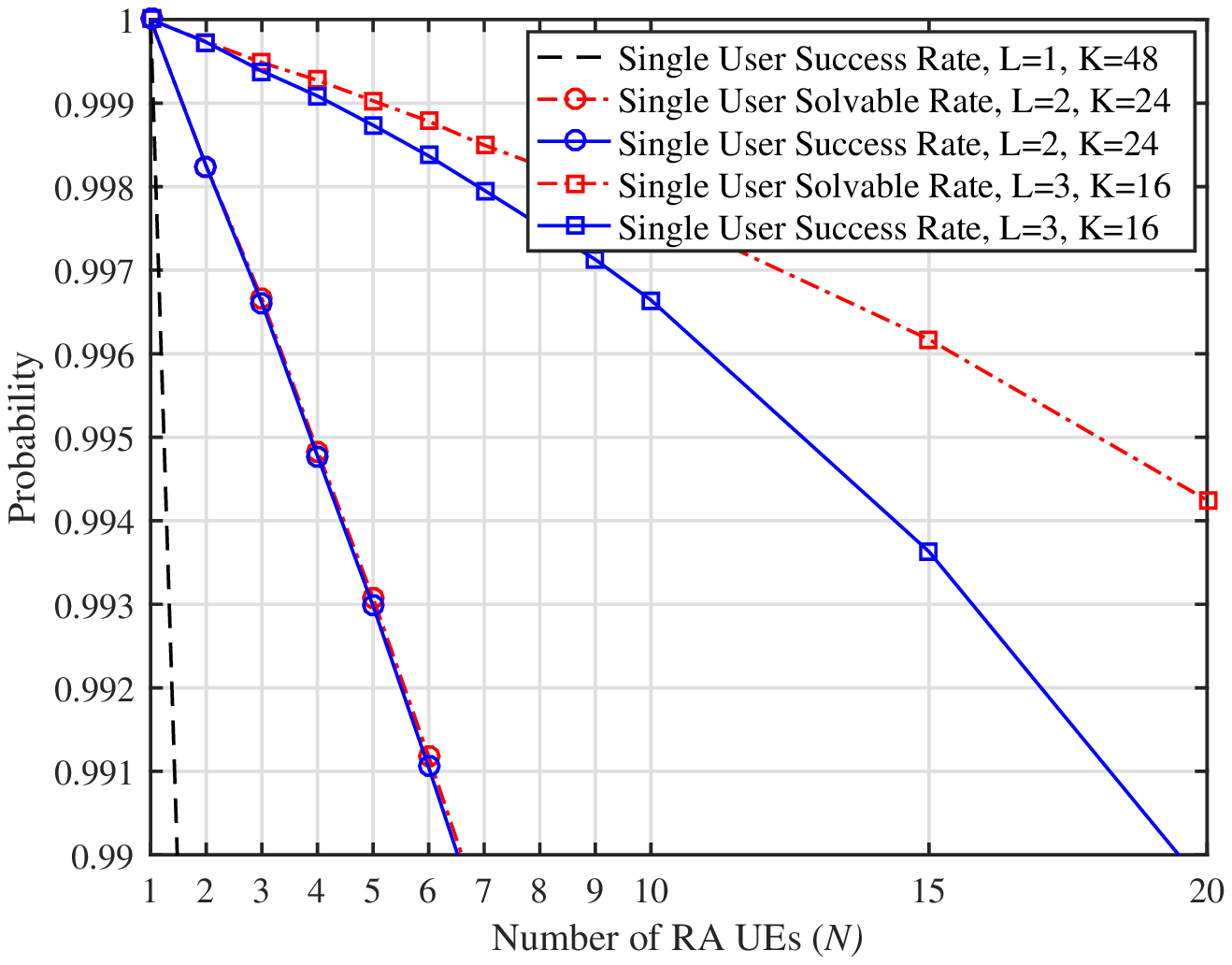}
\caption{$P^{'}_{\rm{success}}$ under spatially correlated fading channel with $M=128$ and $\rm{SNR}=0$ dB.} \label{fig8}
\end{figure}

\begin{figure}[!t]
\centering
\includegraphics[width=3.2in]{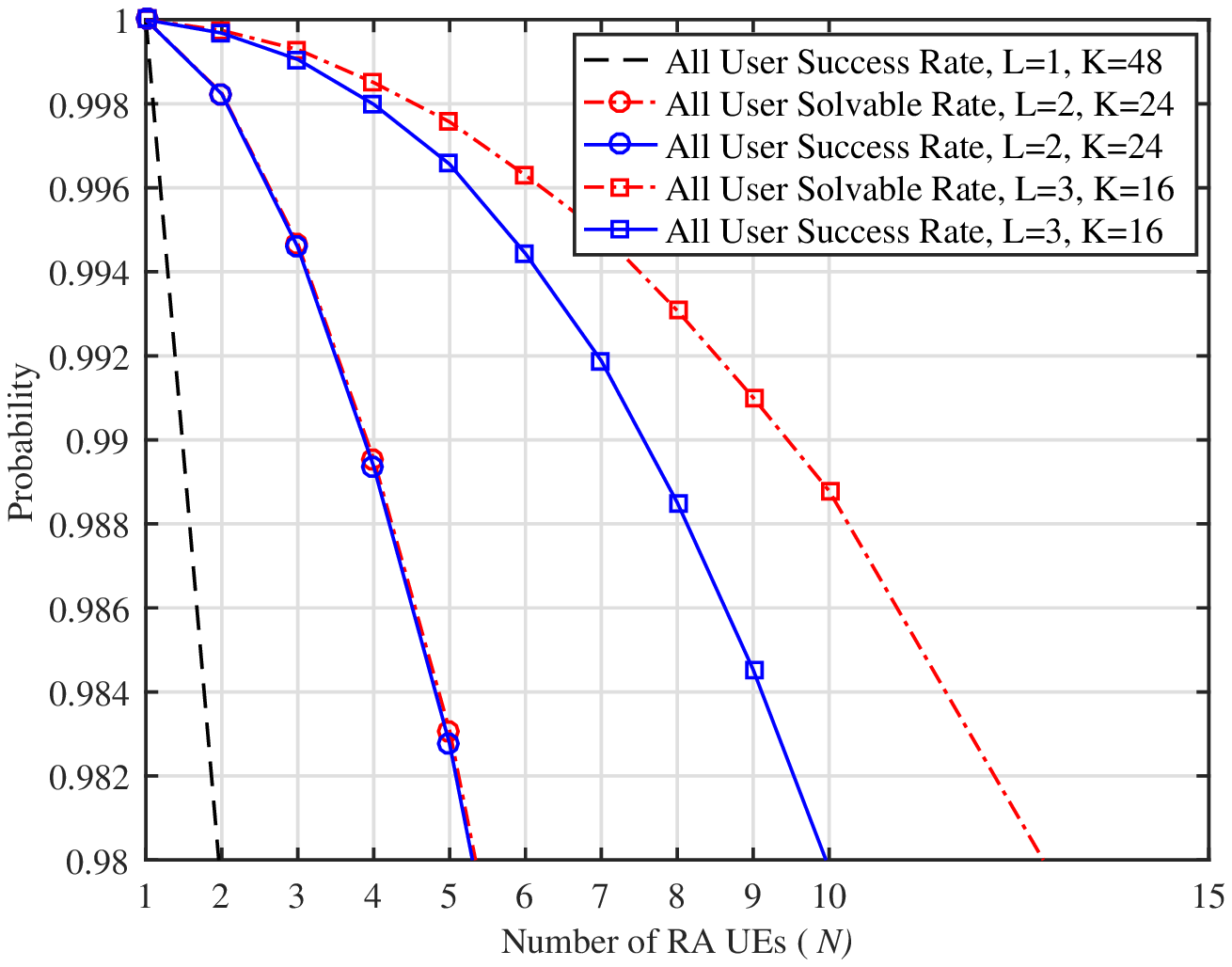}
\caption{$P_{\rm{success}}$ under spatially correlated fading channel with $M=128$ and $\rm{SNR}=0$ dB.} \label{fig6}
\end{figure}

To further evaluate the success rate performance of the proposed multiple preamble approach, a more realistic channel model, i.e., the spatially correlated fading channel, is considered in Fig. \ref{fig8} and \ref{fig6}. Comparing Fig. \ref{fig8} to Fig. \ref{fig7}, it is observed the success rate remains the same when $L=2$ however degrades when $L=3$. The degradation is due to the reason that the increased channel spatial correlations among antennas cause certain loss of the quasi-orthogonality among UEs. Although the UE detection performance of the proposed algorithm decreases under spatially correlated fading channel, we still observe that the number of RA UEs that the BS can serve is as high as 19 at $P^{'}_{\rm{success}}=0.99$, which is more than three times that of $L=2$ and more than ten times that of single preamble. Please be noted that in all these simulations, $KL$ is kept constant, i.e., the total length of preambles are kept constant and the only variation is the number of preambles that we break the total length into. Similar observations are obtained when comparing Fig. \ref{fig6} to Fig. \ref{fig5}.

In conclusion, the proposed UE detection algorithm provides satisfactory performance that enables the high success rate of RA with massive MIMO and super preamble with $L=2$ and $L=3$.

\subsection{NMSE Performance of Channel Estimation}
\begin{figure}[!t]
\centering
\includegraphics[width=3.2in]{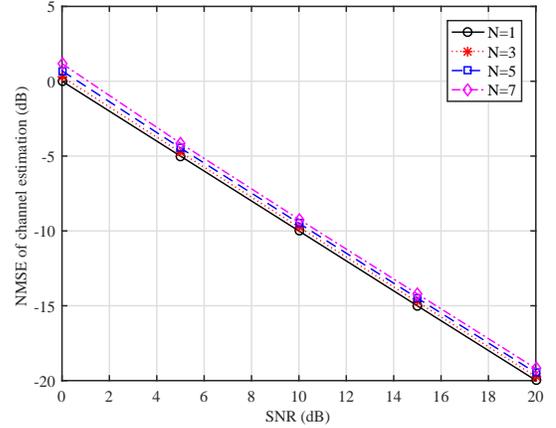}
\caption{Channel estimation performance under spatially correlated fading channel with $L=3$, $K=16$ and $M=128$.} \label{fig10}
\end{figure}

Fig. \ref{fig10} presents the NMSE performance of channel estimation with the super preambles vs. SNR with different number of simultaneous RA UEs under spatially correlated fading channel. We see that the NMSE increases as the number of simultaneous RA UEs $N$ increases, which is due to the fact that the super preambles of RA UEs are not orthogonal in general. Nevertheless, the increase of NMSE is rather slight in average, where it is about 1dB when $N=7$.

\section{Conclusions}
In this paper, a super preamble consisting of $L$ consecutive preambles, along with the UE detection and channel estimation method, is proposed for high success rate of grant-free RA with massive MIMO. We theoretically analyzed the solvable rate of RA UEs with multiple preambles, and simulation results verified the accuracy of the analysis and confirmed that multiple preambles are very effective in increasing solvable rate. It was also shown that the proposed UE detection algorithm provides satisfactory performance that enables the high success rate of RA with massive MIMO and super preamble with $L=2$ and $L=3$. Specifically, turning a preamble into a super preamble consisting of two or three shorter preambles, without increasing preamble resources, the success rate of grant-free RA could be significantly increased, with the help of massive MIMO.

\ifCLASSOPTIONcaptionsoff
  \newpage
\fi

\end{document}